\pdfoutput=1

\documentclass[12pt,a4paper]{article}

\usepackage{lineno}  

\usepackage{graphicx}  

\usepackage{xspace}
\usepackage{color}
\usepackage{colortbl}

\usepackage{amsmath}

\usepackage{ifthen} 

\usepackage{cite}

\newboolean{pdflatex}
\setboolean{pdflatex}{true} 
\newboolean{uprightparticles}
\setboolean{uprightparticles}{false} 
\usepackage{amssymb}
\usepackage{amsfonts}
\usepackage{upgreek}

\usepackage{hyperref}
\usepackage[all]{hypcap} 

\def\lhcb {LHCb\xspace}
\def\ux85 {UX85\xspace}

\def\belle  {Belle\xspace}

\def\cdf    {CDF\xspace}

\ifthenelse{\boolean{uprightparticles}}%
{

 \def\Ppi         {\ensuremath{\uppi}\xspace}

 \def\PDelta      {\ensuremath{\Delta}\xspace}                 
 \def\PXi      {\ensuremath{\Xi}\xspace}                 
 \def\PLambda      {\ensuremath{\Lambda}\xspace}                 
 \def\PSigma      {\ensuremath{\Sigma}\xspace}                 
 \def\POmega      {\ensuremath{\Omega}\xspace}                 
 \def\PUpsilon      {\ensuremath{\Upsilon}\xspace}                 
 
 \def\PB      {\ensuremath{\mathrm{B}}\xspace}                 
                  
 \def\PD      {\ensuremath{\mathrm{D}}\xspace}

 \def\PK      {\ensuremath{\mathrm{K}}\xspace}

 \def\Ph      {\ensuremath{\mathrm{h}}\xspace}                 
 \def\Pi      {\ensuremath{\mathrm{i}}\xspace}

 \def\Ps      {\ensuremath{\mathrm{s}}\xspace}

}
{

 \def\Ppi         {\ensuremath{\pi}\xspace}

 \mathchardef\PDelta="7101
 \mathchardef\PXi="7104
 \mathchardef\PLambda="7103
 \mathchardef\PSigma="7106
 \mathchardef\POmega="710A
 \mathchardef\PUpsilon="7107
                  
 \def\PB      {\ensuremath{B}\xspace}                 
                  
 \def\PD      {\ensuremath{D}\xspace}

 \def\PK      {\ensuremath{K}\xspace}

 \def\Ph      {\ensuremath{h}\xspace}                 
 \def\Pi      {\ensuremath{i}\xspace}

 \def\Ps      {\ensuremath{s}\xspace}

}

\def\squark    {\ensuremath{\Ps}\xspace}

\def\pion  {\ensuremath{\Ppi}\xspace}

\def\pip   {\ensuremath{\pion^+}\xspace}
\def\pim   {\ensuremath{\pion^-}\xspace}

\def\kaon  {\ensuremath{\PK}\xspace}
  \def\Kbar  {\kern 0.2em\overline{\kern -0.2em \PK}{}\xspace}

\def\Kz    {\ensuremath{\kaon^0}\xspace}
\def\Kzb   {\ensuremath{\Kbar^0}\xspace}
\def\KzKzb {\ensuremath{\Kz \kern -0.16em \Kzb}\xspace}
\def\Kp    {\ensuremath{\kaon^+}\xspace}
\def\Km    {\ensuremath{\kaon^-}\xspace}

\def\KpKm  {\ensuremath{\Kp \kern -0.16em \Km}\xspace}
\def\KS    {\ensuremath{\kaon^0_{\rm\scriptscriptstyle S}}\xspace}

  \def\Dbar    {\kern 0.2em\overline{\kern -0.2em \PD}{}\xspace}
\def\D       {\ensuremath{\PD}\xspace}

\def\Dz      {\ensuremath{\D^0}\xspace}
\def\Dzb     {\ensuremath{\Dbar^0}\xspace}
\def\DzDzb   {\ensuremath{\Dz {\kern -0.16em \Dzb}}\xspace}
\def\Dp      {\ensuremath{\D^+}\xspace}
\def\Dm      {\ensuremath{\D^-}\xspace}

\def\DpDm    {\ensuremath{\Dp {\kern -0.16em \Dm}}\xspace}

\def\B       {\ensuremath{\PB}\xspace}
  \def\Bbar    {\kern 0.18em\overline{\kern -0.18em \PB}{}\xspace}

\def\Bz      {\ensuremath{\B^0}\xspace}

\def\Bd      {\ensuremath{\B^0}\xspace}
\def\Bs      {\ensuremath{\B^0_\squark}\xspace}
\def\Bsb     {\ensuremath{\Bbar^0_\squark}\xspace}

  \def\Y#1S{\ensuremath{\PUpsilon{(#1S)}}\xspace}

\def\L {\ensuremath{\PLambda}\xspace}

\newcommand{\decay}[2]{\ensuremath{#1\!\to #2}\xspace}         
\def\ra                 {\ensuremath{\rightarrow}\xspace}
\def\to                 {\ensuremath{\rightarrow}\xspace}

\newcommand{\tauBz}{\ensuremath{\tau_{\Bz}}\xspace}

\def\CP                {\ensuremath{C\!P}\xspace}

\newcommand{\DGs}{\ensuremath{\Delta\Gamma_{\squark}}\xspace}

\newcommand{\Gs}{\ensuremath{\Gamma_{\squark}}\xspace}

\def\BTohh        {\decay{\B}{\Ph^+ \Ph'^-}}

\def\BdToKpi      {\decay{\Bd}{\Kp\pim}}
\def\BsToKK       {\decay{\Bs}{\Kp\Km}}

\def\AT#1     {\ensuremath{A_{\mathrm{T}}^{#1}}\xspace}           

\def\C#1      {\ensuremath{\mathcal{C}_{#1}}\xspace}                       
\def\Cp#1     {\ensuremath{\mathcal{C}_{#1}^{'}}\xspace}                    
\def\Ceff#1   {\ensuremath{\mathcal{C}_{#1}^{\mathrm{(eff)}}}\xspace}        
\def\Cpeff#1  {\ensuremath{\mathcal{C}_{#1}^{'\mathrm{(eff)}}}\xspace}       
\def\Ope#1    {\ensuremath{\mathcal{O}_{#1}}\xspace}                       
\def\Opep#1   {\ensuremath{\mathcal{O}_{#1}^{'}}\xspace}                    

\newcommand{\tev}{\ensuremath{\mathrm{\,Te\kern -0.1em V}}\xspace}
\newcommand{\gev}{\ensuremath{\mathrm{\,Ge\kern -0.1em V}}\xspace}
\newcommand{\mev}{\ensuremath{\mathrm{\,Me\kern -0.1em V}}\xspace}
\newcommand{\kev}{\ensuremath{\mathrm{\,ke\kern -0.1em V}}\xspace}
\newcommand{\ev}{\ensuremath{\mathrm{\,e\kern -0.1em V}}\xspace}
\newcommand{\gevc}{\ensuremath{{\mathrm{\,Ge\kern -0.1em V\!/}c}}\xspace}
\newcommand{\mevc}{\ensuremath{{\mathrm{\,Me\kern -0.1em V\!/}c}}\xspace}
\newcommand{\gevcc}{\ensuremath{{\mathrm{\,Ge\kern -0.1em V\!/}c^2}}\xspace}
\newcommand{\gevgevcccc}{\ensuremath{{\mathrm{\,Ge\kern -0.1em V^2\!/}c^4}}\xspace}
\newcommand{\mevcc}{\ensuremath{{\mathrm{\,Me\kern -0.1em V\!/}c^2}}\xspace}

\def\mm   {\ensuremath{\rm \,mm}\xspace}

\def\pb {\ensuremath{\rm \,pb}\xspace}
\def\invpb {\ensuremath{\mbox{\,pb}^{-1}}\xspace}

\def\invfb   {\ensuremath{\mbox{\,fb}^{-1}}\xspace}

\def\ps   {\ensuremath{{\rm \,ps}}\xspace}
\def\fs   {\ensuremath{\rm \,fs}\xspace}

\newcommand{\stat}{\ensuremath{\mathrm{(stat)}}\xspace}
\newcommand{\syst}{\ensuremath{\mathrm{(syst)}}\xspace}

\def\gsim{{~\raise.15em\hbox{$>$}\kern-.85em
          \lower.35em\hbox{$\sim$}~}\xspace}
\def\lsim{{~\raise.15em\hbox{$<$}\kern-.85em
          \lower.35em\hbox{$\sim$}~}\xspace}

\def\sqs   {\ensuremath{\protect\sqrt{s}}\xspace}

\def\tell1  {TELL1\xspace}
\def\ukl1   {UKL1\xspace}

\newcommand{\etal}{{\slshape et al.\/}\xspace}

\newcommand{\NCBsToKK}{\ensuremath{\B_s \rightarrow KK}\xspace}
\newcommand{\NCBdToKpi}{\ensuremath{\B \rightarrow K \pi}\xspace}

\def\DToKpi            {\decay{\Dz}{\Km\pip}\xspace}

\newcommand{\tauBsToKK} {\ensuremath{\tau_{KK }}\xspace}
\newcommand{\tauBdToKpi}{\ensuremath{\tau_{K\pi}}\xspace}

\newcommand{\RDBsToKK}  {\ensuremath{\tauBsToKK^{-1} ~ - ~ \tauBdToKpi^{-1}}}

\newcommand{\ADGs}{\ensuremath{{\cal A}_{\Delta\Gamma_s}}\xspace}

\def\pdf        {PDF\xspace}

\begin{document}


\pdfoutput=1

\begin{titlepage}
\pagenumbering{roman}

\vspace*{-1.5cm}
\centerline{\large EUROPEAN ORGANIZATION FOR NUCLEAR RESEARCH (CERN)}
\vspace*{1.5cm}
\hspace*{-0.5cm}
\begin{tabular*}{\linewidth}{lc@{\extracolsep{\fill}}r} \\
\ifthenelse{\boolean{pdflatex}}
{\vspace*{-2.7cm}\mbox{\!\!\!\includegraphics[width=.14\textwidth]{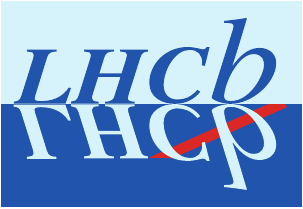}} & &}%
{\vspace*{-1.2cm}\mbox{\!\!\!\includegraphics[width=.12\textwidth]{lhcb-logo.eps}} & &}%
\\
 & & CERN-PH-EP-2011-167 \\  
 & & LHCb-PAPER-2011-014 \\  
 & & \today \\ 
 & & \\
\end{tabular*}

\vspace*{4.0cm}

{\bf\boldmath\huge
\begin{center}
  Measurement of the effective \BsToKK lifetime
\end{center}
}

\vspace*{2.0cm}

\begin{center}
The LHCb Collaboration\footnote{Authors are listed on the following pages.}
\end{center}

\vspace{\fill}

\begin{abstract}
  \noindent
A measurement of the effective \BsToKK lifetime is presented
using approximately 37~\invpb of data collected by \lhcb during
2010. This quantity can be used to put constraints on contributions
from processes beyond the Standard Model in the \Bs meson system and
is determined by two complementary approaches as
\begin{displaymath}
  \tauBsToKK = 1.440\pm0.096~\stat\pm0.008~\syst\pm0.003~(\mathrm{model})~\ps.
\end{displaymath}

\end{abstract}

\vspace*{2.0cm}
\vspace{\fill}

\end{titlepage}


\newpage
\setcounter{page}{2}
\mbox{~}
\newpage

\begin{flushleft}

R.~Aaij$^{23}$, 
C.~Abellan~Beteta$^{35,n}$, 
B.~Adeva$^{36}$, 
M.~Adinolfi$^{42}$, 
C.~Adrover$^{6}$, 
A.~Affolder$^{48}$, 
Z.~Ajaltouni$^{5}$, 
J.~Albrecht$^{37}$, 
F.~Alessio$^{37}$, 
M.~Alexander$^{47}$
% ...

{\it
% ...
$ ^{5}$Clermont Universit\'{e}, Universit\'{e} Blaise Pascal, CNRS/IN2P3, LPC, Clermont-Ferrand, France\\
$ ^{6}$CPPM, Aix-Marseille Universit\'{e}, CNRS/IN2P3, Marseille, France\\
% ...
$ ^{23}$Nikhef National Institute for Subatomic Physics, Amsterdam, Netherlands\\
% ...
$ ^{35}$Universitat de Barcelona, Barcelona, Spain\\
$ ^{36}$Universidad de Santiago de Compostela, Santiago de Compostela, Spain\\
$ ^{37}$European Organization for Nuclear Research (CERN), Geneva, Switzerland\\
% ...
$ ^{42}$H.H. Wills Physics Laboratory, University of Bristol, Bristol, United Kingdom\\
% ...
$ ^{47}$School of Physics and Astronomy, University of Glasgow, Glasgow, United Kingdom\\
$ ^{48}$Oliver Lodge Laboratory, University of Liverpool, Liverpool, United Kingdom\\
% ...
$ ^{n}$LIFAELS, La Salle, Universitat Ramon Llull, Barcelona, Spain\\
% ...
}
\end{flushleft}

\cleardoublepage




\pagestyle{plain} 
\setcounter{page}{1}
\pagenumbering{arabic}


\section{Introduction}
\label{sec:intro}
The study of charmless \PB meson decays of the form $\BTohh$, where
$h^{(\prime)}$ is either a kaon, pion or proton, offers a rich
opportunity to explore the phase structure of the CKM matrix and to
search for manifestations of physics beyond the Standard Model.  The
effective lifetime, defined as the decay time expectation value, of
the \Bs meson measured in the decay channel \BsToKK (charge conjugate
modes are implied throughout the paper) is of considerable interest as
it can be used to put constraints on contributions from new physical
phenomena to the \Bs meson
system~\cite{bib:Grossman,Lenz:2006hd,bib:Fleischer1,bib:Fleischer2}.
The \BsToKK decay was first observed by
\cdf~\cite{bib:CDFBs,bib:CDFBs2}. The decay has subsequently been
confirmed by \belle~\cite{bib:Belle2009}.

The detailed formalism of the effective lifetime in \BsToKK decay can
be found in Refs.~\cite{bib:Fleischer1,bib:Fleischer2}.  The untagged
decay time distribution can be written as
\begin{eqnarray}
  \label{eqn:dblExp}
  \Gamma(t) & \propto & \left ( 1 - \ADGs  \right )e^{-\Gamma_L t} + \left ( 1 + \ADGs \right )e^{-\Gamma_H t} \, .
\end{eqnarray}

The parameter \ADGs is defined as $\ADGs = -2 {\rm
  Re}(\lambda)/\left(1 + |\lambda|^2\right)$ where $\lambda =
(q/p)(\overline{A}/A)$ and the complex coefficients $p$ and $q$ define
the mass eigenstates of the \Bs--\Bsb system in terms of the flavour
eigenstates (see, {\it e.g.}, Ref.~\cite{PDG}), while $A$
($\overline{A}$) gives the amplitude for \Bs (\Bsb) decay to the \CP
even $\Kp\Km$ final state.  In the absence of \CP violation, ${\rm
  Re}(\lambda) = 1$ and $\rm{Im(\lambda)=0}$, so that the distribution
involves only the term containing $\Gamma_L$. Any deviation from a
pure single exponential with decay constant $\Gamma^{-1}_L$ is a
measure of \CP violation.

When modelling the decay time distribution shown in
Eq.~\ref{eqn:dblExp} with a single exponential function in a maximum
likelihood fit, it converges to the effective lifetime given in
Eq.~\ref{eq:tauKKPred}~\cite{bib:Hartkorn99}. For small values of the
relative width difference $\DGs/\Gs = (\Gamma_L - \Gamma_H)/\left(
(\Gamma_L + \Gamma_H)/2\right)$, the distribution can be approximated
by Taylor expansion as shown in the second part of the
equation~\cite{bib:Fleischer1}
\begin{equation}
  \label{eq:tauKKPred}
  \tauBsToKK = \tau_{\Bs} \frac{1}{1-y_s^2} \left [\frac{1 +
      2\ADGs y_s + y_s^2}{1 + \ADGs y_s} \right] = \tau_{\Bs}
  \left( 1 + \ADGs y_s + \mathcal{O}(y_s^2) \right),
\end{equation}

\noindent where $\tau_{\Bs} = 2 / \left ( \Gamma_H + \Gamma_L \right )
= \Gs^{-1}$ and $y_s = \DGs/2\Gs$. The Standard Model predictions for
these parameters are
$\ADGs=-0.97^{+0.014}_{-0.009}$\cite{bib:Fleischer1} and
$y_s=0.066\pm0.016$\cite{Nierste:CKM}.

The decay \BsToKK is dominated by loop diagrams carrying, in the
Standard Model, the same phase as the $\Bs$--$\Bsb$ mixing amplitude
and hence the measured effective lifetime is expected to be close to
$\Gamma_L^{-1}$. The tree contribution to the \BsToKK decay amplitude,
however, introduces \CP violation effects. The Standard Model
prediction is $ \tauBsToKK = 1.390\pm0.032~\ps$~\cite{bib:Fleischer1}.
In the presence of physics beyond the Standard Model, deviations of
the measured value from this prediction are possible.
 
The measurement has been performed using a data sample corresponding
to an integrated luminosity of $37~\invpb$ collected by \lhcb at an
energy of $\sqrt{s} = 7$ TeV during 2010.  A key aspect of the
analysis is the correction of lifetime biasing effects, referred to as
the acceptance, which are introduced by the selection criteria to
enrich the \B meson sample.  Two complementary data-driven approaches
have been developed to compensate for this bias.  One method relies on
extracting the acceptance function from data, and then applies this
acceptance correction to obtain a measurement of the \BsToKK
lifetime. The other approach cancels the acceptance bias by taking the
ratio of the \BsToKK lifetime distribution with that of \BdToKpi.


\section{Data Sample}
\label{sec:selection} 

The \lhcb detector \cite{bib:LHCb} is a single arm spectrometer with a
pseudorapidity acceptance of $2<\eta<5$ for charged particles.  The
detector includes a high precision tracking system which consists of a
silicon vertex detector and several dedicated tracking planes with
silicon microstrip detectors (Inner Tracker) covering the region with
high charged particle multiplicity and straw tube detectors (Outer
Tracker) for the region with lower occupancy. The Inner and Outer
trackers are placed after the dipole magnet to allow the measurement
of the charged particles' momenta as they traverse the detector.
Excellent particle identification capabilities are provided by two
ring imaging Cherenkov detectors which allow charged pions, kaons and
protons to be distinguished from each other in the momentum range
2--100 \gevc. The experiment employs a multi-level trigger to reduce
the readout rate and enhance signal purity: a hardware trigger based
on the measurement of the energy deposited in the calorimeter cells
and the momentum transverse to the beamline of muon candidates, as
well as a software trigger which allows the reconstruction of the full
event information.

\B mesons are produced with an average momentum of around 100 \gevc
and have decay vertices displaced from the primary interaction
vertex. Background particles tend to have low momentum and tend to
originate from the primary $pp$ collision. These features are
exploited in the event selection.  In the absolute lifetime
measurement the final event selection is designed to be more stringent
than the trigger requirements, as this simplifies the calculation of
the candidate's acceptance function. The tracks associated with the
final state particles of the \B meson decay are required to have a
good track fit quality ($\chi^2$/ndf $< 3$ for one of the two tracks
and $\chi^2$/ndf $< 4$ for the other), have high momentum ($p > 13.5$
\gevc), and at least one particle must have a transverse momentum of
more than 2.5 \gevc.  The primary proton-proton interaction vertex (or
vertices in case of multiple interactions) of the event is fitted from
the reconstructed charged particles. The reconstructed trajectory of
at least one of the final state particles is required to have a
distance of closest approach to all primary vertices of at least
0.25\mm.

The \B meson candidate is obtained by reconstructing the vertex formed
by the two-particle final state. The \B meson transverse momentum is
required to be greater than 0.9 \gevc and the distance of the decay
vertex to the closest primary $pp$ interaction vertex has to be larger
than 2.4\mm. In the final stage of the selection the modes \BsToKK and
\BdToKpi are separated by pion/kaon likelihood variables which use
information obtained from the ring imaging Cherenkov detectors.

The event selection used in the relative lifetime analysis is very
similar. However, some selection criteria can be slightly relaxed as
the analysis does not depend on the exact trigger requirements.

\pdfoutput=1

\section{Relative Lifetime Measurement}
\label{sec:relMeth}

\begin{figure}[t]
  \centering
  \includegraphics[width=0.49\textwidth]{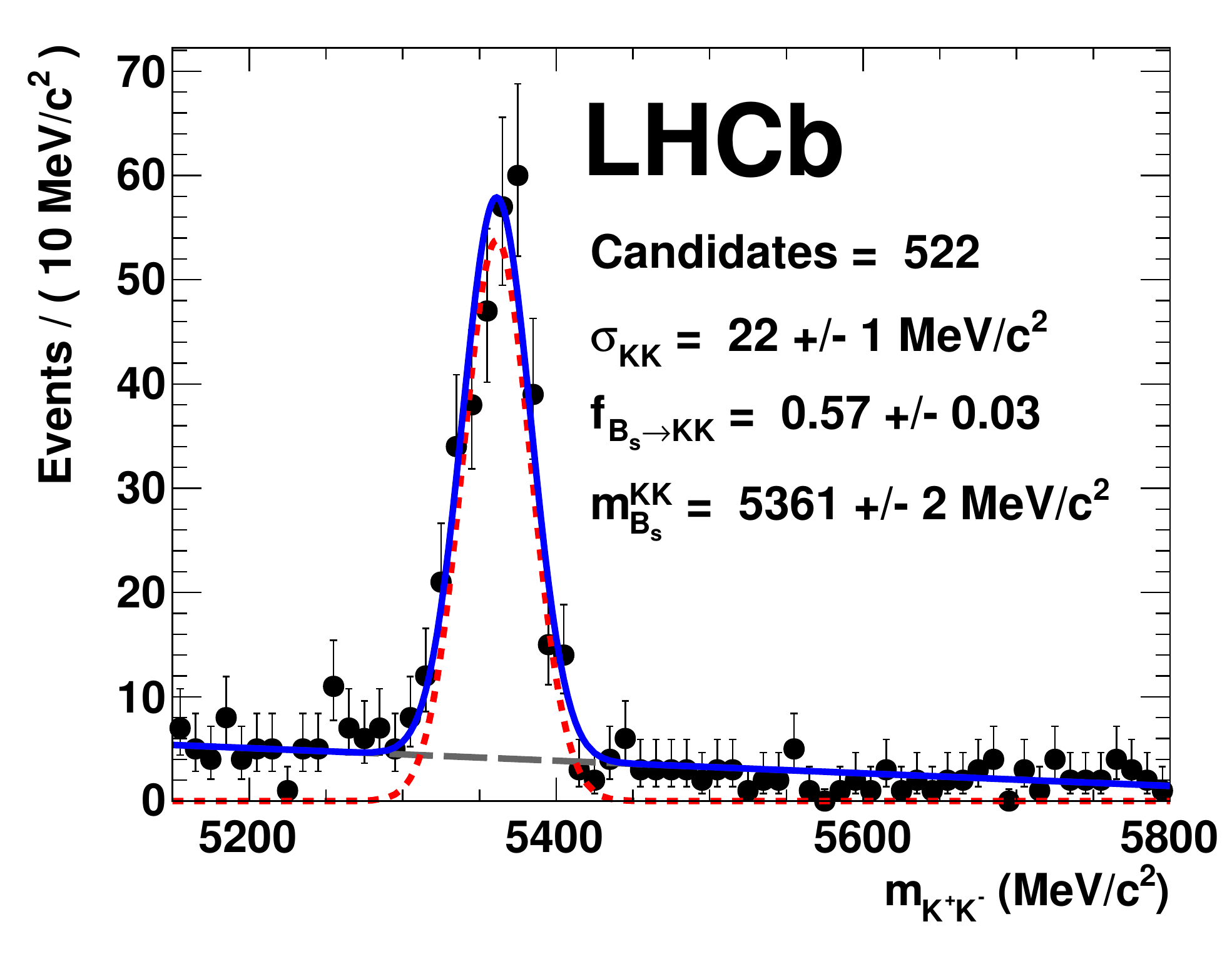}
  \includegraphics[width=0.49\textwidth]{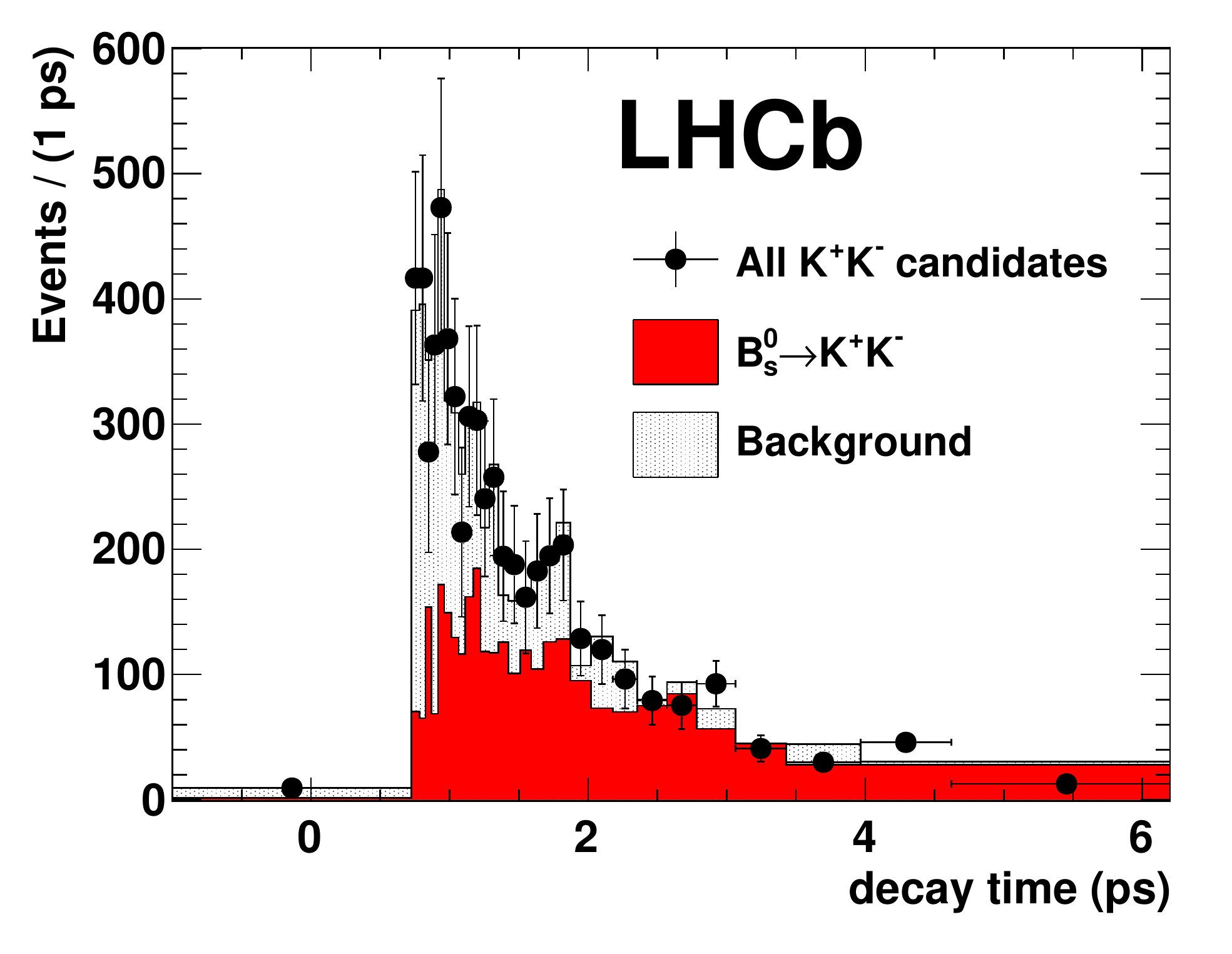}
  \caption{Results of the relative lifetime fit. Left: Fit to the
    time-integrated $KK$ mass spectrum. Right: Fit to the $KK$ decay
    time distribution. The black points show the total number of
    candidates per picosecond in each decay time bin, the stacked
    histogram shows the \BsToKK yield in red (dark) and the background
    yield in grey (light).}
  \label{fig:KKResults}
\end{figure}

This analysis exploits the fact that the kinematic properties of the
\BsToKK decay are very similar to those of \BdToKpi.  The two
different decay modes can be separated using information from the ring
imaging Cherenkov detectors. The left part of Fig.~\ref{fig:KKResults}
shows the invariant mass distribution of the \BsToKK candidates after
the final event selection. In addition 1,424 \BdToKpi candidates are selected.
Using a data-driven particle identification
calibration method described in the systematics section, the remaining
contamination in the \BsToKK sample from other \BTohh final states in
the analysed mass region is estimated to be 3.8\%.

\B mesons in either channel can be selected using identical kinematic
constraints and hence their decay time acceptance functions are almost
identical. Therefore the effects of the decay time acceptance cancel
in the ratio and the effective \BsToKK lifetime can be extracted {\it
  relative} to the \BdToKpi mode from the variation of the ratio
$R(t)$ of the yield of \B meson candidates in both decay modes with
decay time :

\begin{equation}
  \label{eqn:RelRate}
  R(t) = R(0) e^{-t \left(\RDBsToKK \right) }.
\end{equation}

\noindent The cancellation of acceptance effects has been verified
using simulated events, including the full simulation of detector
effects, trigger response and final event selection. Any non-cancelling
acceptance bias on the measured lifetime is found to be smaller 1\fs.

In order to extract the effective \BsToKK lifetime, the yield of \B
meson candidates is determined in bins of decay time for both decay
modes. Thirty bins between -1\ps and 35\ps are chosen such that each
bin contains approximately the same number of \B meson candidates.
The ratio of the yields is then fitted as a function of decay time and
the relative lifetime can be determined according to
Eq.~\ref{eqn:RelRate}. With this approach it is not necessary to
parametrise the decay time distribution of the background. In order to
maximise the statistical precision, both steps of the analysis are
combined in a simultaneous fit to the $\Kp \Km$ and $\Kp \pim$
invariant mass spectra across all decay time bins. The signal
distributions are described by Gaussian functions and the
combinatorial background by first order polynomials. The parameters of
the signal and background probability density functions (PDFs) are
fixed to the results of time-integrated mass fits before the lifetime
fit is performed. The \BdToKpi yield ($N_{\NCBdToKpi}$) is allowed to
float freely in each bin but the \BsToKK yield ($N_{\NCBsToKK}$) is
constrained to follow

\begin{equation}
  N_{\NCBsToKK}(\bar{t}_i) =
  N_{\NCBdToKpi}(\bar{t}_i)R(0)e^{-\bar{t}_i \left(\RDBsToKK\right) },
\end{equation}

\noindent where $\bar{t}_i$ is the mean decay time in the $i^{\rm th}$
bin. In total the simultaneous fit has 94 free parameters and tests
using Toy Monte Carlo simulated data have found the fit to be unbiased
to below 1\fs on the measured \BsToKK lifetime. Each mass fit used in
the simultaneous fit is unbinned and must be split into mass bins in
order to evaluate the fit $\chi^2$. Two mass bins are chosen, one
signal dominated and one background dominated, in order to guarantee a
minimum of 5-6 candidates in each bin.  Using this appraoch the
$\chi^2$ per degree of freedom of the simultaneous fit is found to be
0.82. The right part of Fig. \ref{fig:KKResults} shows the decay time
distribution obtained from the fit and the fitted reciprocal lifetime
difference is

\begin{displaymath}
  \RDBsToKK = 0.013 ~ \pm ~ 0.045 ~ \stat ~ \ps^{-1}.
\end{displaymath}

\noindent Taking the \BdToKpi lifetime as equal to the mean \Bz
lifetime ($\tauBz = 1.519 \pm 0.007~\ps$) \cite{PDG}, this measurement
can be expressed as

\begin{displaymath}
 \tauBsToKK = 1.490\pm0.100~\stat\pm0.007~\mbox{(input)~\ps}
\end{displaymath}

\noindent where the second uncertainty originates from the uncertainty
of the \Bz lifetime.

\pdfoutput=1
\section{Absolute Lifetime Measurement}
\label{sec:absMeth}

\begin{figure}[!!!htbp]
  \centering
   \begin{minipage}{0.42\textwidth}
    (a)\\
    \includegraphics[width=0.9\textwidth]{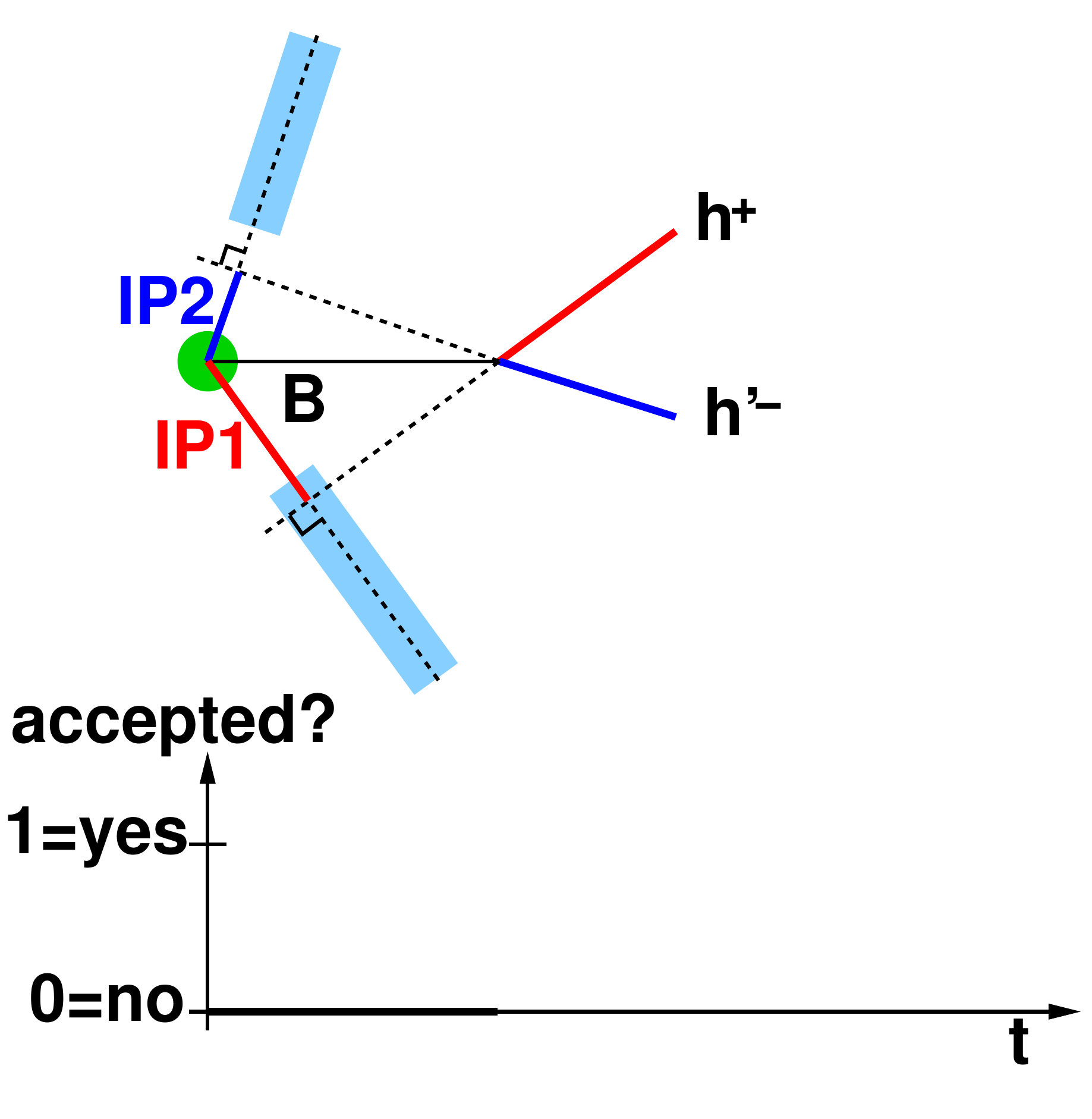}
  \end{minipage}
  \begin{minipage}{0.42\textwidth}
    (b)\\
    \includegraphics[width=0.87\textwidth]{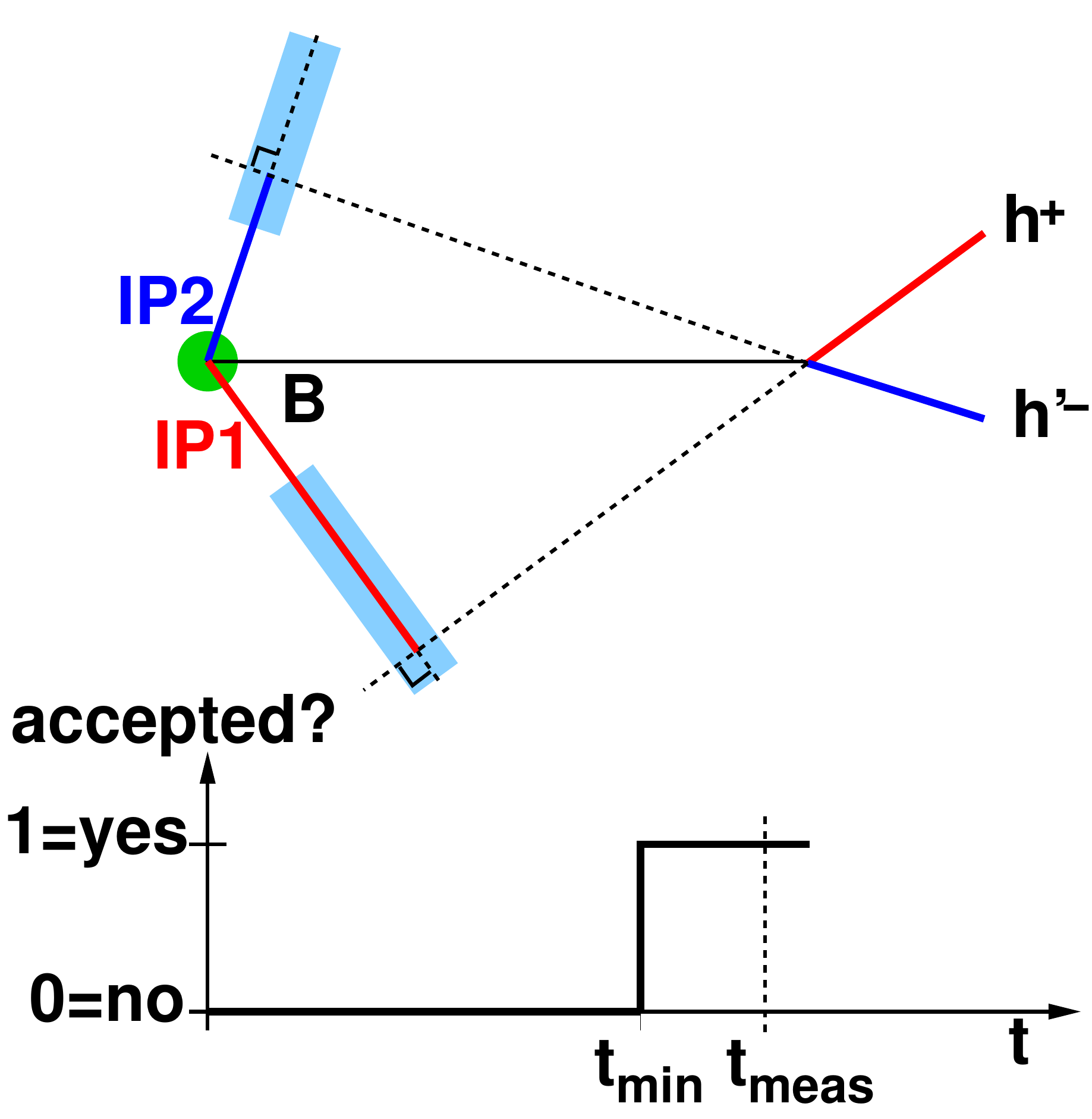}
  \end{minipage}
  \vspace{10mm}\\
  \caption[Decay-time acceptance function for an event with a two-body
    hadronic decay.]{Decay-time acceptance function for an event of a
    two-body hadronic decay. The light blue (shaded) regions show the
    bands for accepting the impact parameter of a track. The impact
    parameter of the negative track (IP2) is too small in (a) and lies
    within the accepted range in (b). The actual measured decay time
    lies in the accepted region.  The acceptance intervals give
    conditional likelihoods used in the lifetime fit.}
  \label{fig:fit_swimming}
\end{figure}

The absolute lifetime measurement method directly determines the
effective \BsToKK lifetime using an acceptance correction calculated
from the data.  This method was first used at the NA11 spectrometer at
CERN SPS~\cite{Bailey:1985zz}, further developed within
\cdf~\cite{Rademacker:2005ay,Aaltonen:2010ta} and was subsequently
studied and implemented in
\lhcb~\cite{Gligorov:2008zza,bib:marco_thesis}.  The {\it per event}
acceptance function is determined by evaluating whether the candidate
would be selected for different values of the \B meson candidate decay
time. For example, for a \B meson candidate, with given kinematic
properties, the measured decay time of the \B meson candidate is
directly related to the point of closest approach of the final state
particles to the associated primary vertex.  Thus a selection
requirement on this quantity directly translates into a discrete
decision about acceptance or rejection of a candidate as a function of
its decay time. This is illustrated in Fig.~\ref{fig:fit_swimming}. In
the presence of several reconstructed primary interaction vertices,
the meson may enter a decay-time region where one of the final state
particles no longer fulfills the selection criteria with respect to
another primary vertex. Hence the acceptance function is determined as
a series of step changes. These \emph{turning points} at which the
candidates enter or leave the acceptance of a given primary vertex
form the basis of extracting the per-event acceptance function in the
data.  The turning points are determined by moving the reconstructed
primary vertex position of the event along the \B meson momentum
vector, and then reapplying the event selection criteria.  The
analysis presented in this paper only includes events with a single
turning point.  The drop of the acceptance to zero when the final
state particles are so far downstream that one is outside the detector
acceptance occurs only after many lifetimes and hence is safely
neglected.

\begin{figure}[htbp]
  \begin{center} 
  \includegraphics*[width=0.49\textwidth]{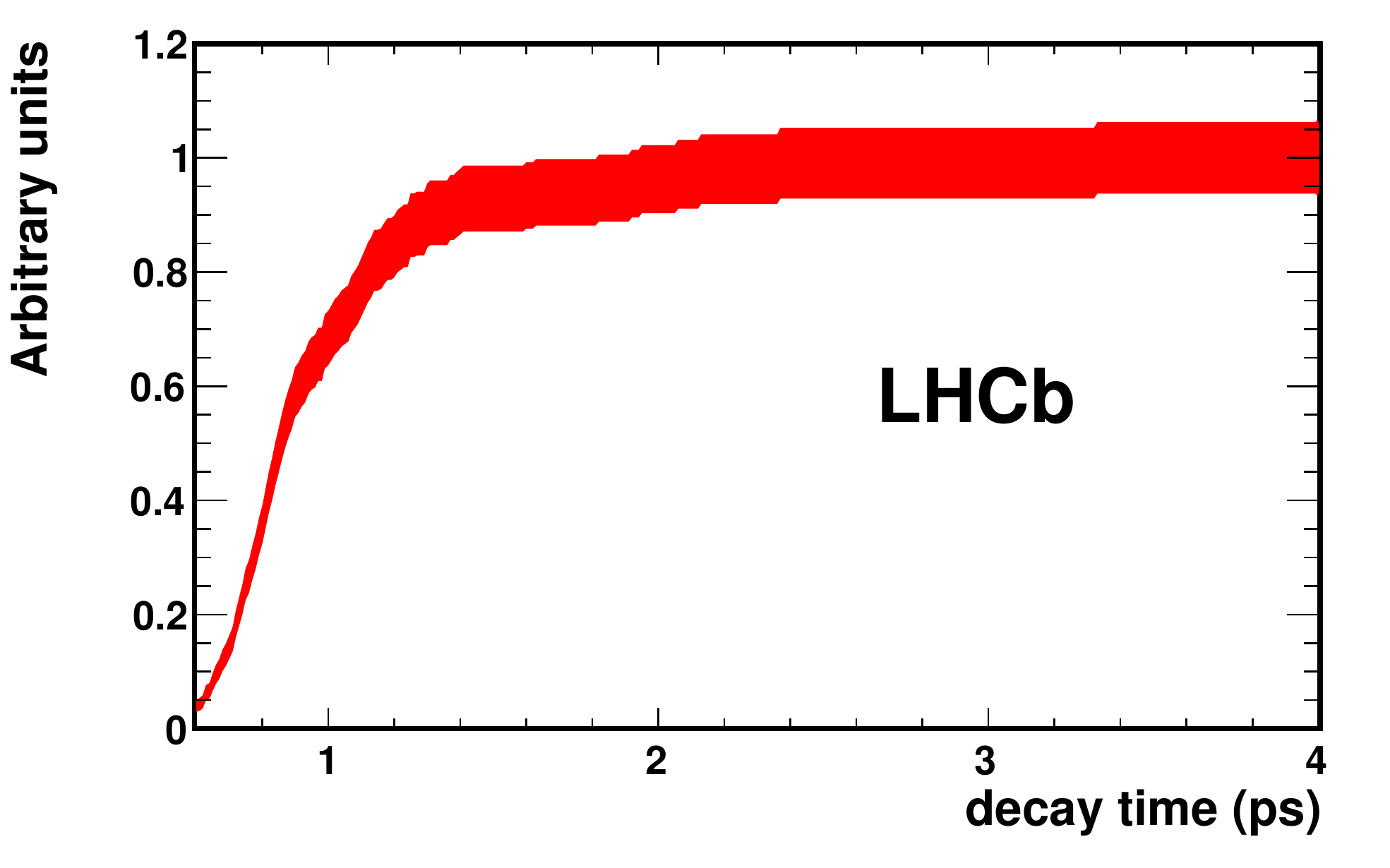} 
  \includegraphics*[width=0.49\textwidth]{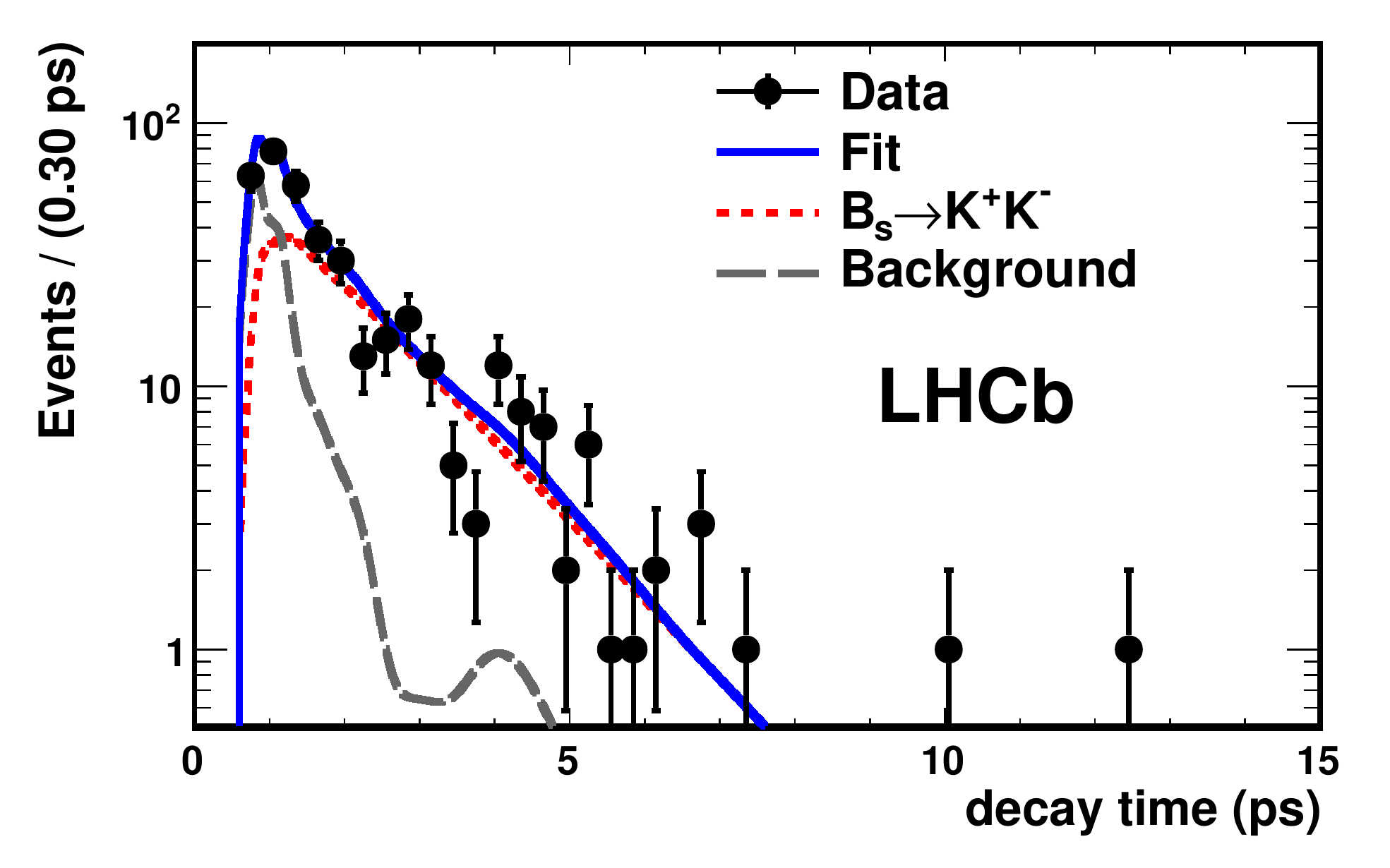} 
  \end{center} 
  \caption[\BsToKK average acceptance function and lifetime
    fit.]{\small Left: Average decay-time acceptance function for
    signal events, where the error band is an estimate of the
    statistical uncertainty. The plot is scaled to 1 at large decay
    times, not taking into account the total signal efficiency.
    Right: Decay-time distribution of the \BsToKK candidates and the
    fitted functions. The estimation of the background distribution is
    sensitive to fluctuations due to the limited statistics. Both
    plots are for the absolute lifetime
    measurement.} \label{fig:absFit}
\end{figure}

The distributions of the turning points, combined with the decay-time
distributions, are converted into an average acceptance function (see
Fig.~\ref{fig:absFit}). The average acceptance is not used in the
lifetime fit, except in the determination of the background decay-time
distribution.

The effective \BsToKK lifetime is extracted by an unbinned maximum
likelihood fit using an analytical probability density function (\pdf)
for the signal decay time and a non-parametric \pdf for the
combinatorial background, as described below. The measurement is
factorised into two independent fits.

A first fit is performed to the observed mass spectrum and used to
determine the signal and background probabilities of each
event. Events with \Bs candidates in the mass range $5272-5800$ \mevcc
were used, hence reducing the contribution of partially reconstructed
background and contamination of \Bd decays below the \Bs mass
peak. The signal distribution is modelled with a Gaussian, and the
background with a linear distribution. The fitted mass value is
compatible with the current world average~\cite{PDG}.

The signal and background probabilities are used in the subsequent
lifetime fit. The decay-time \pdf of the signal is calculated
analytically taking into account the per-event acceptance and the
decay-time resolution. The decay-time \pdf of the combinatorial
background is estimated from data using a non-parametric method and is
modelled by a sum of kernel functions which represent each candidate
by a normalised Gaussian function centred at the measured decay time
with a width proportional to an estimate of the density of candidates
at this decay time~\cite{bib:cranmer}. The lifetime fit is performed
in the decay-time range of $0.6-15~\ps$, hence only candidates within
this range were accepted. The analysis was tested on the \BdToKpi
channel, for which a lifetime compatible with the world average value
was obtained, and applied to the \BsToKK channel only once the full
analysis procedure had been fixed. The result of the lifetime fit is

\begin{displaymath}
\tauBsToKK = 1.440\pm0.096~\stat \; \mathrm{ps}
\end{displaymath}
and is illustrated in Fig.~\ref{fig:absFit}.

\section{Systematic Uncertainties}
\label{sec:syst}

\begin{table}[ht]
\begin{center}
\caption{\small Summary of systematic uncertainties on the \BsToKK lifetime measurements.} 
\begin{minipage}[t]{0.99\textwidth}
  \centering
  \begin{tabular}{|l|l|l|}
    \hline
    Source of uncertainty               & Uncertainty on            & Uncertainty on        \\ 
                                        & \tauBsToKK (fs)           & \RDBsToKK (ns$^{-1}$) \\ \hline \hline
    Fit method                          & 3.2                       &   \\ 
    Acceptance correction               & 6.3                       &  $\quad$ 0.5 \\
    Mass model                          & 1.9                       &   \\
    \BTohh background                   & 1.9                       & 1.4    \\
    Partially reconstructed background  & 1.9                       & 1.1   \\
    Combinatorial background            & 1.5                       & 1.6   \\
    Primary vertex association          & 1.2                       & 0.5  \\   
    Detector length scale               & 1.5                       & 0.7    \\
    Production asymmetry                & 1.4                       & 0.6 \\
    Minimum accepted lifetime           & 1.1                       & N/A \\ \hline \hline
    Total (added in quadrature)         & 8.4                       & 2.7 \\ \hline
    Effective lifetime interpretation   & 2.8                       & 1.1 \\ \hline
    \end{tabular}
\label{tab:syst}
\end{minipage}
\begin{minipage}[t]{0.01\textwidth}
\vspace{-6.2cm}$\qquad \qquad \qquad  \qquad \Bigg\}$
\end{minipage}
\end{center}
\end{table}

The systematic uncertainties are listed in Table~\ref{tab:syst} and
discussed below. The dominant contributions to the systematic
uncertainty for the absolute lifetime measurement come from the
treatment of the acceptance correction ($6.3~\fs$) and the fitting
procedure ($3.2~\fs$). The systematic uncertainty from the acceptance
correction is determined by applying the same analysis technique to a
kinematically similar high statistics decay in the charm sector
(\DToKpi \cite{charmConfNote}). This analysis yields a lifetime value
in good agreement with the current world average and of better
statistical accuracy. The uncertainty on the comparison between the
measured value and the world average is rescaled by the \B meson and
charm meson lifetime ratio.  The uncertainty due to the fitting
procedure is evaluated using simplified simulations. A large number of
pseudo-experiments are simulated and the pull of the fitted lifetimes
compared to the input value to the fit is used to estimate the
accuracy of the fit. These sources of uncertainty are not dominant in
the relative method, and are estimated from simplified simulations
which also include the systematic uncertainty of the mass model. Hence
a common systematic uncertainty is assigned to these three sources.

The effect of the contamination of other \BTohh modes to the signal
modes is determined by a data-driven method.  The misidentification
probability of protons, pions and kaons is measured in data using the
decays $\KS \ra \pip \pim$, $\Dz \ra \Kp \pim$, $\phi \ra \Kp \Km$ and
$\L \ra p \pim$, where the particle type is inferred from kinematic
constraints alone \cite{bib:Powell2010}. As the particle
identification likelihood separating protons, kaons and pions depends
on kinematic properties such as momentum, transverse momentum, and
number of reconstructed primary interaction vertices, the sample is
reweighted to reflect the different kinematic range of the final state
particles in \BTohh decays.  The effect on the measured lifetime is
evaluated with simplified simulations.

Decays of \Bs and \Bd to three or more final state particles, which
have been partially reconstructed, lie predominantly in the mass range
below the \Bs mass peak outside the analysed region. Residual
background from this source is estimated from data and evaluated with
a sample of fully simulated partially reconstructed decays.  The
effect on the fitted lifetime is then evaluated.

In the absolute lifetime measurement, the combinatorial background of
the decay time distribution is described by a non-parametric function,
based on the observed events with masses above the \Bs meson signal
region. The systematic uncertainty is evaluated by varying the region
used for evaluating the combinatorial background.  In the relative
lifetime measurement, the combinatorial background in the $hh^\prime$
invariant mass spectrum is described by a first order polynomial. To
estimate the systematic uncertainty, a sample of simulated events is
obtained with a simplified simulation using an exponential function,
and subsequently fitted with a first order polynomial.

Events may contain several primary interactions and a reconstructed
\PB meson candidate may be associated to the wrong primary
vertex. This effect is studied using the more abundant charm meson
decays where the lifetime is measured separately for events with only
one or any number of primary vertices and the observed variation is
scaled to the \PB meson system.

Particle decay times are measured from the distance between the
primary vertex and secondary decay vertex in the silicon vertex
detector.  The systematic uncertainty from this source is determined
by considering the potential error on the length scale of the detector
from the mechanical survey, thermal expansion and the current
alignment precision.

The analysis assumes that \Bs and \Bsb mesons are produced in equal
quantities. The influence of a production asymmetry for \Bs mesons on
the measured lifetime is found to be small.

In the absolute lifetime method both a Gaussian and a Crystal Ball
mass model \cite{bib:CB} are implemented and the effect on fully
simulated data is evaluated to estimate the systematic uncertainty due
to the modelling of the signal \pdf. In the relative lifetime method
this uncertainty is evaluated with simplified simulations and included
in the fitting procedure uncertainty.

In the absolute \BsToKK lifetime measurement a cut is applied on the
minimal reconstructed decay time. As the background decay time
estimation will smear this step in the distribution, a systematic
uncertainty is quoted from varying this cut.

There is an additional uncertainty introduced if the result is
interpreted using Eq.~\ref{eq:tauKKPred}, as this expression does not
take into account detector resolution and decay time acceptance. This
effect was studied using simplified simulations modelling the
acceptance observed in the data and conservative values of \DGs = 0.1
\ps and \ADGs = -0.6.  The observed bias with respect to the
prediction of Eq.~\ref{eq:tauKKPred} is 3 \fs. This effect is labelled
``Effective lifetime interpretation'' in Table~\ref{tab:syst} and is
not a source of systematic uncertainty on the measurement but is
relevant to the interpretation of the measured lifetime.

\section{Results and Conclusions}
\label{sec:conclusions}

The effective \BsToKK lifetime has been measured in $pp$ interactions
using a data sample corresponding to an integrated luminosity of
$37\pb^{-1}$ recorded by the \lhcb experiment in 2010. Two
complementary approaches have been followed to compensate for
acceptance effects introduced by the trigger and final event selection
used to enrich the sample of \Bs mesons.  The absolute measurement
extracts the {\it per event} acceptance function directly from the
data and finds:
\begin{displaymath}
  \tauBsToKK = 1.440\pm0.096~\stat\pm0.008~\syst\pm0.003~(\mathrm{model})~\ps
\end{displaymath}

\noindent
where the third source of uncertainty labelled ``model'' is related to
the interpretation of the effective lifetime.

The relative method exploits the fact that the kinematic properties of
the various \BTohh modes are almost identical and extracts the \BsToKK
lifetime relative to the \BdToKpi lifetime as:
\begin{displaymath}
  \RDBsToKK = 0.013\pm0.045~\stat\pm0.003~\syst\pm0.001~(\mathrm{model})~\ps^{-1}.
\end{displaymath}
Taking the \BdToKpi lifetime as equal to the mean \Bz lifetime ($\tauBz = 1.519 \pm 0.007~\ps$)
\cite{PDG}, this measurement can be expressed as:
\begin{displaymath}
 \tauBsToKK = 1.490\pm0.100~\stat\pm0.006~\syst\pm0.002~(\mathrm{model})\pm0.007~\mbox{(input)~\ps}.
\end{displaymath}
where the last uncertainty originates from the uncertainty of the \Bz
lifetime.  Both measurements are found to be compatible with each
other, taking the overlap in the data analysed into account.

Due to the large overlap of the data analysed by the two methods and
the high correlation of the systematic uncertainties, there is no
significant gain from a combination of the two numbers. Instead, the
result obtained using the absolute lifetime method is taken as the
final result. The measured effective \BsToKK lifetime is in agreement
with the Standard Model prediction of $\tauBsToKK =
1.390\pm0.032~\ps$~\cite{bib:Fleischer1}.

\section*{Acknowledgements}

\noindent We express our gratitude to our colleagues in the CERN
accelerator departments for the excellent performance of the LHC. We
thank the technical and administrative staff at CERN and at the LHCb
institutes, and acknowledge support from the National Agencies: CAPES,
CNPq, FAPERJ and FINEP (Brazil); CERN; NSFC (China); CNRS/IN2P3
(France); BMBF, DFG, HGF and MPG (Germany); SFI (Ireland); INFN
(Italy); FOM and NWO (The Netherlands); SCSR (Poland); ANCS (Romania);
MinES of Russia and Rosatom (Russia); MICINN, XuntaGal and GENCAT
(Spain); SNSF and SER (Switzerland); NAS Ukraine (Ukraine); STFC
(United Kingdom); NSF (USA). We also acknowledge the support received
from the ERC under FP7 and the Region Auvergne.



\bibliographystyle{model1-num-names}
\bibliography{<your-bib-database>}



\end{document}